\documentstyle [prl,aps,epsfig,floats]{revtex}

\makeatletter           
\@floatstrue
\def\figure{\let\@capwidth\columnwidth\@float{figure}}
\let\endfigure\end@float
\@namedef{figure*}{\let\@capwidth\textwidth\@dblfloat{figure}}
\@namedef{endfigure*}{\end@dblfloat}
\def\table{\let\@capwidth\columnwidth\@float{table}}
\let\endtable\end@float
\@namedef{table*}{\let\@capwidth\textwidth\@dblfloat{table}}
\@namedef{endtable*}{\end@dblfloat}

\makeatother

\tightenlines
\begin {document}

\twocolumn[\hsize\textwidth\columnwidth\hsize\csname
@twocolumnfalse\endcsname


\preprint {UW/PT-01-07}
\title {
   Relativistic Turbulence:
   A Long Way from Preheating to Equilibrium
}

\author {Raphael Micha}

\address
    {%
    Theoretische Physik, 
    ETH Z\"urich, 
    CH-8093 Z\"urich, 
    Switzerland 
    }%

\author {Igor I. Tkachev}
\address
    {%
      Theory Division, CERN, CH-1211 Geneva 23, Switzerland \\
      and \\
      Institute for Nuclear Research of the Russian Academy of
      Sciences, 117312, Moscow, Russia
    }%

\maketitle

\begin {abstract}%
{%
  We study, both numerically and analytically, the development of
  equilibrium after preheating.  We show that the process is
  characterised by the appearance of Kolmogorov spectra and the
  evolution towards thermal equilibrium follows self-similar
  dynamics.  Simplified kinetic theory gives values for all
  characteristic exponents which are close to what is observed in
  lattice simulations. The resulting time for thermalization is long, and
  temperature at thermalization is low, $T \sim 100$ eV in the simple
  $\lambda \Phi^4$ inflationary model. Our results allow a
  straightforward generalization to realistic models.
}%
\end {abstract}

\vskip2pc]


\paragraph*{Introduction.}
The dynamics of equilibration and thermalization of field theories is
of interest for various reasons.  In high-energy physics understanding
of these processes is crucial for applications to heavy ion collisions
and to reheating of the early universe after inflation.  Inflation
solves the flatness and the horizon problems of the standard big bang
cosmology and provides a calculable mechanism by which initial density
perturbations were generated \cite{books}.  At the end of inflation
the Universe was in a vacuum-like state.  In the process of decay of
this state and subsequent thermalization (reheating) the matter
content of the universe is created. It was realized recently that the
initial stage of reheating, dubbed preheating
\cite{preheat}, is a fast, explosive process. This initial stage by
now is well understood
\cite{Khlebnikov:1996mc,Prokopec:1996rr,Kofman:1997yn,hybrid,Giudice:1999fb}.
Strong and fast amplification of fluctuation fields at low momenta may
lead to various interesting physical effects, like non-thermal phase
transitions \cite{Kofman:1995fi}, peculiar baryogenesis \cite{baryo},
generation of high-frequency gravitational waves
\cite{Khlebnikov:1997di}, etc.

Understanding of the subsequent stages of reheating and thermalization
processes and calculation of the final equilibrium temperature is
important for various applications, most notably baryogenesis and the
problem of over-abundant gravitino production in supergravity models
\cite{gravitino}.  Thermalization of field theories was discussed
already, see e.g. Refs. \cite{thermalization}.  However, at present
the process of thermalization after preheating is still far away from
being well understood and developed. The problem is that at the
preheating stage the occupation numbers are very large, of order of
the inverse coupling constant. In addition, in many models the zero
mode does not decay completely. Therefore, a simple kinetic approach
is not applicable.

Fortunately, the description in terms of classical field theory is
valid in this situation \cite{Khlebnikov:1996mc}, and the process of
preheating, as well as subsequent thermalization, can be studied on a
lattice.  In this paper we adopt this approach. Our goal is to
integrate the system on a lattice sufficiently accurately and
sufficiently far in time to be able to see generic features, and
possibly to the stage, at which the kinetic description becomes a  good
approximation scheme. Lattice studies of thermalization, similar to
ours, were done in Ref. \cite{Felder:2000hr}. Several generic rules of
thermalization were formulated, like the early equipartition of energy
between coupled fields.  However, the problem is very complicated and
there are other unanswered important questions like what is the
final thermalization temperature, at what stage the kinetic
description becomes valid, what is the functional form of particle
distributions during the thermalization stage, etc.

For our study we use a higher accuracy, improved version of the
LATTICEEASY code \cite{Felder:2000hq}. We show that the distribution
functions follow a {\em self-similar} evolution related to the turbulent
transport of wave energy. This property enables us to estimate the physical
reheating temperature, which turns out to be very low.  The concept
should be rather model independent since typical ranges of particle
momenta at preheating and in thermal equilibrium are widely separated.
However, in this letter we will restrict our numerical integration
(but not the discussion) to the ``minimal'' inflationary model, the
massless $\lambda \Phi^4$-theory.

\paragraph*{The Model.}
With conformal coupling to gravity and after a  rescaling of the field,
$\varphi \equiv  \Phi a$, where $a(t)$ is the cosmological scale
factor, the equation of motion in comoving coordinates describes a
$\varphi^4$-theory in Minkowski space-time,
\begin{equation}
\Box \varphi + \lambda \varphi^3 = 0.
\label{eqn_mot}
\end{equation}
At the end of inflation the field is homogeneous, 
$\varphi = \varphi_0 (t)$.  Later on fluctuations develop,
but the homogeneous component of the field, which corresponds to the zero
momentum in the Fourier decomposition, may be dynamically important and is
referred to as the ``zero-mode.''  In such situations it is convenient
to make a further rescaling of the field, $\phi \equiv
\varphi/\varphi_0(t_0)$, and of the space-time coordinates, $x^\mu
\rightarrow \sqrt{\lambda} \varphi_0(t_0) x^\mu$, which transforms the
equation of motion (\ref{eqn_mot}) into dimensionless and parameter
free form,
\begin{equation}
\Box \phi + \phi^3 = 0  \,  .
\label{BEq}
\end{equation}
Here $t_0$ corresponds to the initial moment of time (end of
inflation), and in what follows we denote dimensionless time as
$\tau$.  With this rescaling the initial condition for the zero-mode
oscillations is $\phi_0(\tau_0 ) = 1 $.  All model dependence on the
coupling constant $\lambda$ and on the initial amplitude of the field
oscillations now is encoded  in the initial conditions for the small
(vacuum) fluctuations of the field with non-zero momenta \cite{Khlebnikov:1996mc}.
The physical normalization of the inflationary model corresponds to
a dimensionful initial amplitude of $\varphi_0(t_0) \approx 0.3 M_{\rm
  Pl}$ and a coupling constant $\lambda \sim 10^{-13}$ \cite{books}.
The re-parametrization property of the system allows to chose a larger
value of $\lambda$ for numerical simulations. We have used
$\lambda=10^{-8}$.

\paragraph*{Numerical Procedure and Results.}
We use a 3-D cubic lattice with periodic boundary conditions.
The finite-differences scheme that was used is 2nd order in time and 4-th
order in space.  The results displayed here are taken from a
simulation with $256^3$ sites and a physical box size $L=14\pi$.  With
this box size the infrared modes which belong to the resonance band
are still well represented, while the ultraviolet lattice cut-off is
sufficiently far away from the occupied modes, such that the particle spectra are not distorted even
at late times.  We have studied the dependence of our results on the
lattice- and the box size to avoid lattice artifacts.  Various quantities
are measured and monitored both in configuration space (zero mode,
$\phi_0 \equiv \langle\phi\rangle$, and the variance, $var(\phi) \equiv
\langle\phi^2\rangle -\phi_0^2$) and in the Fourier space.
Using fourier transformed fields we first define the wave amplitudes
(which correspond to annihilation operators in the quantum problem),
\begin{equation}
a({\vec k}) \equiv \frac{\omega_k \phi_{\vec k} +i \dot{\phi}_{\vec k}}
{{(2\pi)^{3/2}}\sqrt{2\omega_k}} \; .
\end{equation}
The effective frequency $\omega_k \equiv \sqrt{k_{\rm D}
  ^2+m_{eff}^2}$ is determined by the effective mass
$m_{eff}^2=3\lambda\langle\phi^2\rangle$ and the inverse $k^2_{\rm D}$
of the lattice Laplacian. In our numerical scheme the latter is given
by
\begin {equation}      
   k_{\rm D} ^2 = b^{-2} \sum_{i \in \{1,2,3\} }\left(
    \frac{5}{2} - \frac{8}{3}
    \cos(b k_i) + \frac{1}{6} \cos(2 b k_i) \right).\label{inv_laplace}
\end {equation}
Here $b=2\pi/L=1/7$ is the lattice constant. 
Making use of $a_k$, we calculate various correlators, 
$n(k) \equiv \langle a^\dagger
a\rangle$, $\sigma (k) \equiv \langle a a\rangle$, $\langle a^\dagger a^\dagger a
a\rangle$, etc. The first one, which corresponds to the particle
occupation numbers, is of prime interest.
 
\begin{figure}
  \includegraphics[width=3.8in]{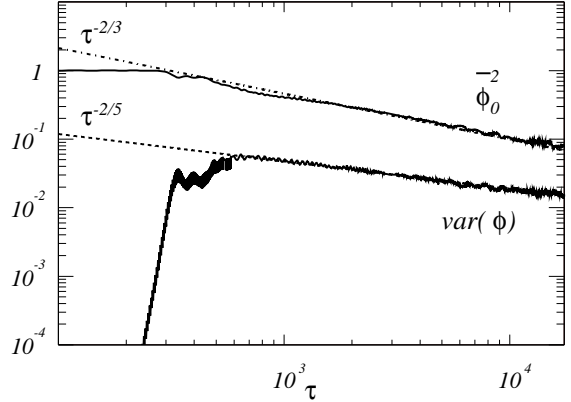}
  \caption{
    Amplitude of the zero-mode oscillations, 
    $\overline\phi_0^2$, and variance of the field fluctuations
    as functions of time $\tau$.
    }
  \label{spat_av}
\end{figure}

We begin the discussion of our numerical results with the evolution of
the zero-mode and the variance of the field, which are shown in
Fig.~\ref{spat_av}.
Initially we see an exponentially fast transfer of the zero-mode
energy into fluctuations during preheating (up to $\tau\sim 300$).  It is
followed by a long and slow relaxation phase. In this regime
($\tau>1500$) the amplitude of the zero mode oscillations 
decreases according to $\sim \tau^{-1/3}$, the variance of the field
(averaged over high-frequency oscillations) drops according to $\sim
\tau^{-2/5}$. This is consistent with previous results
\cite{Khlebnikov:1996mc}.  In addition we find  that in this regime the
zero-mode is in a non-trivial dynamical equilibrium with the bath of
highly occupied modes: when the zero-mode is artificially removed, it
is recreated on a short time-scale (Bose condensation).

\begin{figure}
  \includegraphics[width=3.8in]{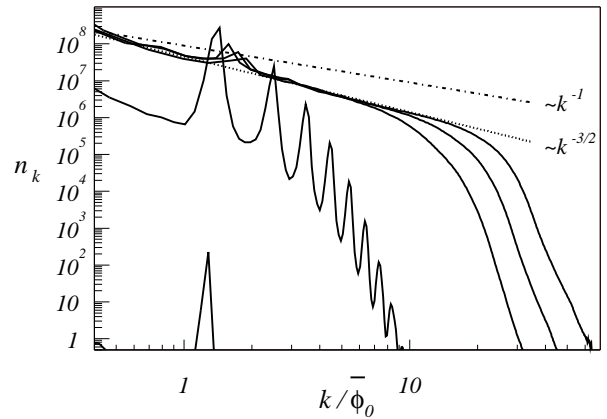}
  \caption{Occupation numbers as  function
  of $k\overline\phi_0^{-1}$ at  $\tau = 100, 400, 2500, 5000, 10000$. }
\label{spectra}
\end{figure}

At early times  the distribution functions of particles over momenta, see Fig.~\ref{spectra}, 
have peaky structure. The first peak which
corresponds to the parametric resonance is initially at the
theoretically predicted value of $k \sim 1.27$
\cite{Kofman:1997yn}. Later ($\tau>1500$) the spectra become smooth and
at small $k$ approach a power-law, $n_k\sim k^{-s}$, where $s$
fluctuates in the range of $1.5 - 1.7$, depending on time and the
range of $k$ where it is fitted. This power law clearly differs from
the the classical thermal equilibrium, $n_k\sim \omega_k^{-1}$. 
It is followed by the exponential cut-off, whose position
monotenously shifts with time towards higher $k$.  Pumping of energy from the
zero mode stays effective all the times (note a small bump in the particle
distributions in Fig.~\ref{spectra} at $k \sim 1$). It corresponds to
the annihilation of four condensed particles into two quanta.
Rescattering of two particles into two particles is also effective.
One of the two can belong to the zero-mode condensate either in
initial or in the final two particle state.  We also can see in
Fig.~\ref{spectra} that in the power-law region $n_k$ is a function of
$k/\overline\phi_0$ only, where $\overline\phi_0(\tau)$ represents
the amplitude of the zero-mode at time $\tau$. This effect can
be related to the above described dynamical equilibrium between
zero-mode and the bath of particles. Indeed, going from
Eq.~(\ref{eqn_mot}) to Eq. (\ref{BEq}) we can rescale by the current
amplitude of the zero-mode.

\begin{figure}
  \includegraphics[width=3.8in]{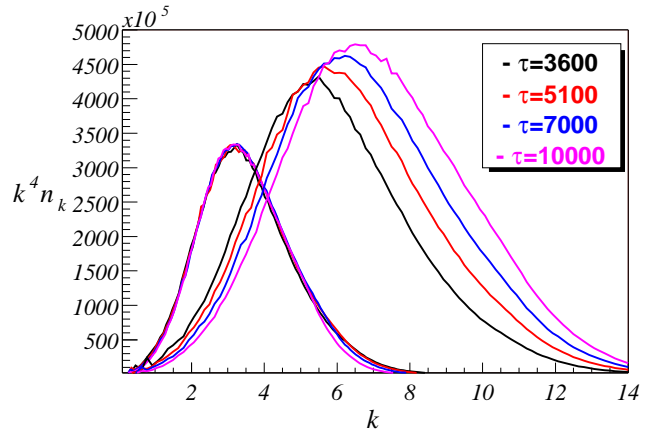} 
  \caption{On the right hand side we plot
    the wave energy per decade found in lattice integration. On the
    left hand side are the same graphs transformed according to
    the relation inverse to Eq.~(\ref{SelfS}).}
\label{self_sim}
\end{figure}

The picture presented in Fig.~\ref{spectra} at late times 
resembles stationary Kolmogorov
turbulence. It appears as such due to rescaling of momenta by the
amplitude of the zero-mode, but in fact in the present model the
turbulence can not be stationary because the amplitude of the
zero-mode (i.e. the strength of the source of turbulence) decreases.
Further examination of Fig~\ref{spectra} suggests that the evolution
of particle spectra may be self-similar. We have tried therefore the
following anzatz
\begin{equation}
n(k,\tau) = \tau^{-q} n_0 (k \tau^{-p}) \, .
\label{SelfS}
\end{equation}
Spectra rescaled at several moments of time by the relation inverse to
Eq.~(\ref{SelfS}) are shown in Fig.~\ref{self_sim}. We have found that
the evolution is indeed self-similar with $q \approx 3.5p$ and $p \approx
1/5$.

\paragraph*{Discussion.}
Here we discuss the question whether a simple kinetic theory gives
predictions for turbulence and self-similarity exponents in agreement
with the lattice calculations. Our lattice study of higher order
correlators, like $\langle a^\dagger a^\dagger a a\rangle$, shows that
the field distribution is very close to Gaussian, see also 
\cite{thermalization,Felder:2000hr}. This facilitates the
use of the kinetic approach. On the other hand we have found that the
magnitude of $\sigma(k)$ is of order of a few percent compared to
$n(k)$, and it is even larger in the region of resonant momenta. This
means that the strict kinetic approach should include $\sigma(k)$.
Nevertheless we neglect these effects and write a kinetic equation in a
simple form $\dot{n_k} = I_k\label{kin1}$, where the collision
integral for a $m$-particle interaction is given by
\begin{equation}
I_k = \int d\Omega_k\, U_k\, F[n] \, .
\label{kin_coll_Int}
\end{equation}
In $d$ spatial dimensions the integration measure $d\Omega_k$ is given
by $m-1$ integrations over $d$-dimensional Fourier space. We include
in it the energy-momentum conservation $\delta$-functions.  But we
do not include there the relativistic $1/\omega (k_i)$ ``on-shell''
factors, which instead appear in the ``matrix element'' of the
corresponding process, $U_k$. This will make the discussion of
relativistic and non-relativistic cases  uniform. The function
$F[n]$ is a sum of products of the type
$n_{k_{j}}^{-1}\prod_{i=1}^{m}n_{k_i}$, where $j \in\{1,\dots,m\}$
with appropriate signs and permutations of indices for incoming and
outgoing particles.  All dynamical aspects of turbulence follow from
the scaling properties of the system \cite{Turb}.  Let
$\omega_k, n_k$ and $U_k$ have defined weights under a $\xi$-rescaling
of Fourier-space,
\begin{eqnarray}
\omega(\xi k_i)&=& \xi^{\alpha}\omega (k_i) \; , \nonumber \\
U(\xi k_1, \dots, \xi k_m) &=& \xi^{\beta} U(k_1, \dots, k_m)\; , \nonumber \\
n(\xi k_i) &=& \xi^{\gamma} n(k_i) \; . \label{pow_law} 
\label{scalins}
\end{eqnarray}
The weight of the full collision integral under this re-parametrization is
\begin{equation}
I_{\xi k}=\xi^{d(m-2)-\alpha+\beta+(m-1)\gamma }I_{k} \, .
\end{equation}
It follows that the stationary turbulence with constant energy
flux over momentum space is characterised by a
power-law distribution function, $n_k\sim k^{-s}$, 
where $s=d+\beta/({m-1})$. The scaling properties also give the
exponents of the self-similar distribution, Eq. (\ref{SelfS}).
Assuming energy conservation in particles and with $\xi = \tau^{-p}$
we find $q = 4p$ and $p = 1/((m-1)\alpha-\beta)$. For stationary turbulence we find that $p$ should be $(m-1)$ times larger.

For a $\lambda \phi^4$-theory in three spatial dimensions and 
four-particle interaction  we have $m=4$, $\beta=-4\alpha$ and $\alpha
= 1$.  In this case $s = 5/3$ and $p = 1/7$. 
For three-particle interaction (the fourth particle belongs to the
condensate in this case and the matrix element contains an additional
factor of $\overline\phi_0^2$) we find $s = 3/2$ and a smaller value
for $p$ compared to the previous case. We can not distinguish between
$5/3$ and $3/2$ for s in our numerical integrations, $s$ rather
fluctuates between these two numbers, while $1/7$ for  $p$ gives
a fit to the data not as good as displayed in Fig.~\ref{self_sim}.
However, during the integration time the energy in particles is
neither conserving, nor there is a stationary source of energy. Namely,
starting from the time at which the solution becomes self-similar, $\tau \sim
3000$, to the end of our integration, the energy influx from the zero
mode to particles is about 20 \%. Correcting for this energy influx
we find $q \approx 3.5p$ and $p \approx 1/6$. 
This should be considered as satisfactory agreement given the 
simplifications which were made.

\paragraph*{Equilibration time and temperature.}
At late times the influence of the zero-mode should become
negligible, but we still may expect the self-similar character of the
evolution. Solution Eq.~(\ref{SelfS}) with $p=1/7$ should be valid in
this case. This allows us to find the time  needed to reach
equilibrium. Indeed, the classical evolution will continue until the
occupation numbers in the region of the peak in Fig.~\ref{self_sim} will
become of order one. At this time quantum effects become important
and the distribution relaxes to thermal. Values of momenta were this
happens are $k_{\rm max} \sim \lambda^{1/4}\varphi_0(\tau_0)$. 
On the other hand the initial distribution is centred around 
$k_0\sim \lambda^{1/2}\varphi_0(\tau_0)$ and moves to ultraviolet
according to Eq.~(\ref{SelfS}) as $\propto k_0 \tau^p$. It follows
that the time to reach equilibrium is
$\tau\sim\lambda^{-7/4}\sim 10^{23}$, where in the second equality we
assumed the normalization to the inflationary model.
For the reheating temperature we find, rotating back from the conformal
reference frame, $T_R\sim k_{max}/a(\tau) \sim \lambda^2 
\varphi_0(t_0)\sim 10^{-26}M_{\rm Pl}\sim 100$ eV, where
for the conformal scale factor we have used $a(\tau) = \tau$.
\paragraph*{Conclusions.}
Reheating after preheating appears to be a rather slow process.
Although the ``effective temperature'' measured at low momentum modes
during preheating may be high, in the model we have considered
the resulting true temperature is parametrically the same as what
could have been obtained in ``naive'' perturbation theory. Namely,
equating the rate of scattering in thermal equilibrium to the Hubble
expansion rate one obtains $T \sim \lambda^2 M_{\rm Pl}$ in this
model.  We anticipate this result should be applicable to more
realistic models of inflation. Note that realistic models involve
many fields and interactions and larger coupling constants will
determine the true temperature.

\paragraph*{Acknowledgements.}
We thank A. Riotto, C. Schmid and D. Semikoz for many useful
discussions during various stages of this project. R.M. thanks the Tomalla
Foundation for financial support.


\begin {references}



\bibitem{books}
A.D.~Linde,
{\it Particle Physics and Inflationary Cosmology},
(Harwood Academic,\ New York,\ 1990);
E.~W.~Kolb\ and\ M.~S.~Turner,
{\em The Early Universe},\ (Addison-Wesley,\ Reading,\ Ma.,\ 1990);
A.R.~Liddle,\ D.H.~Lyth,\                                                       
{\it Cosmological Inflation and Large-Scale Structure},                         
(Cambridge University Press,\ Cambridge,\ 2000 );   
D.~H.~Lyth and A.~Riotto,
Phys.\ Rept.\  {\bf 314}, 1 (1999).


\bibitem{preheat}
L.~Kofman, A.~D.~Linde and A.~A.~Starobinsky,
Phys.\ Rev.\ Lett.\  {\bf 73}, 3195 (1994);
Y.~Shtanov, J.~Traschen and R.~H.~Brandenberger,
Phys.\ Rev.\ D {\bf 51}, 5438 (1995).

\bibitem{Khlebnikov:1996mc}
S.~Y.~Khlebnikov and I.~I.~Tkachev,
Phys.\ Rev.\ Lett.\  {\bf 77}, 219 (1996);
{\it ibid}
{\bf 79}, 1607 (1997);
Phys.\ Lett.\ B {\bf 390}, 80 (1997).

\bibitem{Prokopec:1996rr}
T.~Prokopec and T.~G.~Roos,
Phys.\ Rev.\ D {\bf 55}, 3768 (1997).

\bibitem{Kofman:1997yn}
L.~Kofman, A.~D.~Linde and A.~A.~Starobinsky,
Phys.\ Rev.\ D {\bf 56}, 3258 (1997);
P.~B.~Greene {\it at al}, 
Phys.\ Rev.\ D {\bf 56}, 6175 (1997).

\bibitem{hybrid}
J.~Garcia-Bellido and A.~D.~Linde,
Phys.\ Rev.\ D {\bf 57}, 6075 (1998); 
R.~Micha and M.~G.~Schmidt,
Eur.\ Phys.\ J.\ C {\bf 14}, 547 (2000);
G.~N.~Felder {\it et al},
Phys.\ Rev.\ Lett.\  {\bf 87}, 011601 (2001).

\bibitem{Giudice:1999fb}
G.~F.~Giudice {\it et al}, 
JHEP {\bf 9908}, 014 (1999).


\bibitem{Kofman:1995fi}
L.~Kofman, A.~D.~Linde and A.~A.~Starobinsky,
Phys.\ Rev.\ Lett.\  {\bf 76}, 1011 (1996);
I.~I.~Tkachev,
Phys.\ Lett.\ B {\bf 376}, 35 (1996);
S.~Khlebnikov {\it et al}, 
Phys.\ Rev.\ Lett.\  {\bf 81}, 2012 (1998).




\bibitem{baryo}
E.~W.~Kolb, A.~D.~Linde and A.~Riotto,
Phys.\ Rev.\ Lett.\  {\bf 77}, 4290 (1996);
E.~W.~Kolb, A.~Riotto and I.~I.~Tkachev,
Phys.\ Lett.\ B {\bf 423}, 348 (1998);
J.~Garcia-Bellido {\it et al}, 
Phys.\ Rev.\ D {\bf 60}, 123504 (1999);
J.~Garcia-Bellido and E.~Ruiz Morales,
Phys.\ Lett.\ B {\bf 536}, 193 (2002).



\bibitem{Khlebnikov:1997di}
S.~Y.~Khlebnikov and I.~I.~Tkachev,
Phys.\ Rev.\ D {\bf 56}, 653 (1997).


\bibitem{gravitino}
J.~R.~Ellis, A.~D.~Linde and D.~V.~Nanopoulos,
Phys.\ Lett.\ B {\bf 118}, 59 (1982);
J.~R.~Ellis, D.~V.~Nanopoulos and S.~Sarkar,
Nucl.\ Phys.\ B {\bf 259}, 175 (1985).


\bibitem{thermalization}
D.~T.~Son,
Phys.\ Rev.\ D {\bf 54}, 3745 (1996);
D.~V.~Semikoz,
Helv.\ Phys.\ Acta {\bf 69}, 207 (1996);
G.~Aarts, G.~F.~Bonini and C.~Wetterich,
Nucl.\ Phys.\ B {\bf 587}, 403 (2000);
S.~Davidson and S.~Sarkar,
JHEP {\bf 0011}, 012 (2000);
M.~Salle, J.~Smit and J.~C.~Vink,
Phys.\ Rev.\ D {\bf 64}, 025016 (2001);
E.~Calzetta and M.~Thibeault,
Phys.\ Rev.\ D {\bf 63}, 103507 (2001);
M.~Grana and E.~Calzetta,
Phys.\ Rev.\ D {\bf 65}, 063522 (2002);
G.~Aarts {\it et al},
Phys.\ Rev.\ D {\bf 66}, 045008 (2002);
J.~Berges and J.~Serreau,
hep-ph/0208070.



\bibitem{Felder:2000hr}
G.~N.~Felder and L.~Kofman,
Phys.\ Rev.\ D {\bf 63}, 103503 (2001).


\bibitem{Felder:2000hq}
G.~N.~Felder and I.~Tkachev, LATTICEEASY code,
arXiv:hep-ph/0011159.


\bibitem {Turb}
  A. K. Kolmogorov,
  Dokl.\ Akad.\ Nauk.\ SSSR\ {\bf 30}, 9 (1941);
  A.~V.~Kats, 
  Zh.\ Eksp.\ Teor.\ Fiz. {\bf 71}, 2104-2112 (1976);
  V. E. Zakharov, S. L. Musher, A. M. Rubenchik,
  Phys. Rep. {\bf 129}, 285 (1985);
  G. E. Falkovich and A. V. Shafarenko, J. Nonlinear Sci.
  {\bf 1}, 457 (1991);
  V.~Zakharov, V.~L'vov, G.~Falkovich, {\it Kolmogorov Spectra
  of Turbulence}, Wave turbulence. Springer-Verlag 1992.


\end {references}
\end {document}